\documentclass[conference]{IEEEtran}
\usepackage{cite}
\usepackage{amsmath,amssymb,amsfonts}
\usepackage{algorithmic}
\usepackage{graphicx}
\usepackage{textcomp}
\usepackage{xcolor}

\usepackage{booktabs}
\usepackage{graphicx}
\usepackage{subcaption}
\usepackage{array}
\usepackage{tabularx}

\makeatletter
\newcommand{\linebreakand}{%
  \end{@IEEEauthorhalign}
  \hfill\mbox{}\par
  \mbox{}\hfill\begin{@IEEEauthorhalign}
}
\makeatother

\def\BibTeX{{\rm B\kern-.05em{\sc i\kern-.025em b}\kern-.08em
    T\kern-.1667em\lower.7ex\hbox{E}\kern-.125emX}}

\usepackage{fancyhdr}
\begin{document}

\title{Switchboard-Affect: Emotion Perception Labels from Conversational Speech}

\author{\IEEEauthorblockN{Amrit Romana}
\IEEEauthorblockA{\textit{Apple} \\
Cupertino, USA \\
ak\_romana@apple.com}
\and
\IEEEauthorblockN{Jaya Narain}
\IEEEauthorblockA{\textit{Apple} \\
Cupertino, USA \\
jnarain@apple.com}
\and
\IEEEauthorblockN{Tien Dung Tran}
\IEEEauthorblockA{\textit{Apple} \\
Cupertino, USA \\
dung\_tran@apple.com}
\linebreakand
\IEEEauthorblockN{Andrea Davis}
\IEEEauthorblockA{\textit{Apple} \\
Cupertino, USA \\
andrea\_k\_davis@apple.com}
\and
\IEEEauthorblockN{Jason Fong}
\IEEEauthorblockA{\textit{Apple} \\
Cupertino, USA \\
jason\_fong@apple.com}
\and
\IEEEauthorblockN{Ramya Rasipuram}
\IEEEauthorblockA{\textit{Apple} \\
Cupertino, USA \\
rrasipuram@apple.com}
\and
\IEEEauthorblockN{Vikramjit Mitra}
\IEEEauthorblockA{\textit{Apple} \\
Cupertino, USA \\
vmitra@apple.com}}


\maketitle
\thispagestyle{fancy}

\begin{abstract}
Understanding the nuances of speech emotion dataset curation and labeling is essential for assessing speech emotion recognition (SER) model potential in real-world applications. Most training and evaluation datasets contain acted or pseudo-acted speech (e.g., podcast speech) in which emotion expressions may be exaggerated or otherwise intentionally modified. Furthermore, datasets labeled based on crowd perception often lack transparency regarding the guidelines given to annotators. These factors make it difficult to understand model performance and pinpoint necessary areas for improvement. To address this gap, we identified the Switchboard corpus as a promising source of naturalistic conversational speech, and we trained a crowd to label the dataset for categorical emotions (anger, contempt, disgust, fear, sadness, surprise, happiness, tenderness, calmness, and neutral) and dimensional attributes (activation, valence, and dominance). We refer to this label set as Switchboard-Affect (SWB-Affect). In this work, we present our approach in detail, including the definitions provided to annotators and an analysis of the lexical and paralinguistic cues that may have played a role in their perception. In addition, we evaluate state-of-the-art SER models, and we find variable performance across the emotion categories with especially poor generalization for anger. These findings underscore the importance of evaluation with datasets that capture natural affective variations in speech. We release the labels for SWB-Affect to enable further analysis in this domain. 

\end{abstract}



\section{Introduction}

Speech emotion recognition (SER) has the potential to enhance human-computer interaction, improve our ability to monitor mental health and well-being \cite{cummins2015review, provost2024emotion}, and better understand customer service, entertainment, and education experiences \cite{han2020ordinal, litman2004predicting}. However, our understanding of SER

\vspace{1em}
\hrule
\vspace{1em}
\noindent{\small \copyright 2025 IEEE. Personal use of this material is permitted. Permission from IEEE must be obtained for all other uses, in any current or future media, including reprinting/republishing this material for advertising or promotional purposes, creating new collective works, for resale or redistribution to servers or lists, or reuse of any copyrighted component of this work in other works.

\noindent DOI: 979-8-3315-8015-5}

\noindent models is limited by training and evaluation data, which often does not include spontaneous speech. In this work, we release emotion perception labels for the Switchboard corpus, a widely analyzed dataset of conversations \cite{godfrey1992switchboard}.

A barrier in SER is the lack of labeled naturalistic conversational speech datasets. The majority of datasets contain recordings of actors portraying emotions. IEMOCAP includes actors performing emotional scripts and improvising scenarios \cite{busso2008iemocap}. CREMA-D \cite{cao2014crema}, RAVDESS \cite{livingstone2018ryerson}, and MEAD \cite{wang2020mead} capture actors performing lexically matched sentences with a variety of target emotions. 
While these datasets allow for understanding of emotion expression under controlled conditions, models trained on this data do not generalize to emotion expressions in-the-wild. This is in part because there are differences between spontaneous and acted speech, including in the lexical content \cite{busso2008scripted} and paralinguistics \cite{wilting2006real}, but also because emotion expression in spontaneous speech tends to be more subtle  \cite{schuller2011recognising}.

Lotfian et. al introduced MSP-Podcast \cite{lotfian2017building} to address the need for non-acted speech affect datasets. They mined potentially emotional audio from existing podcast recordings, and trained a crowd to label the speech segments for categorical \cite{ekman1992argument} and dimensional \cite{grimm2006segregated} attributes. Similarly, Zadeh et al. introduced CMU-MOSEI \cite{bagher-zadeh-etal-2018-multimodal} in which they mined segments from YouTube and crowd-sourced emotion perception labels. 
While podcast and YouTube speech contains characteristics closer to spontaneous speech, content creators are often intending to convey a pre-determined story, message, or point of view, and in doing so, they may alter their speaking style and expressions to appear more engaging \cite{martikainen2022exploring}. Due to these intentional expression modifications, we can consider podcast and YouTube speech to be pseudo-spontaneous \cite{tucker2023spontaneous} and evaluation with only those domains may not provide a true representation of SER model performance in-the-wild. 

A small body of work has started to explore emotion expression and perception in more naturalistic conversational settings. Namely, Mooriyad et al. explored emotion in Fisher English Training, an LDC conversational dataset \cite{mariooryad2014building}. They crowd-sourced categorical and dimensional annotations from multiple annotators per segment. They found over half of the random sentences selected contained emotional traits but the labeled data were not released for further analysis. Later, Lu et al. investigated emotion in Switchboard, another LDC conversational dataset \cite{lu2020speech}. They released Switchboard-Sentiment with positive, neutral, and negative labels, but these categories do not capture the full range of emotion labels relevant to broader SER applications. Furthermore, the annotation protocol lacked transparency, including whether multiple annotators were involved and how they were instructed to label the data. Kossaifi et al. released SEWA DB, in which participants watched adverts selected to elicit emotions and discussed the advert with their conversation partner \cite{kossaifi2019sewa}. While these segments were annotated with multiple graders and arousal and valence attributes, they were not annotated for categorical emotions which have been increasingly of interest to the community \cite{das2024speechverse}. In this work, we address the need for emotion perception labels, both categorical and dimensional, pertaining to naturalistic conversational speech.

We introduce affect annotations for Switchboard. Our contributions include: 
\begin{itemize}
    \item The release of affect annotations, including categorical labels (anger, contempt, disgust, fear, sadness, surprise, happiness, tenderness, calmness, and neutral) and dimensional labels (activation, valence, and dominance)\footnote{https://github.com/apple/ml-switchboard-affect}
    \item {An overview of the training process used to prepare graders and an analysis of data trends and distributions}
    \item An investigation into the lexical and paralinguistic cues corresponding to emotion perception in the dataset
    \item An analysis of state-of-the-art SER model performance on the newly annotated data 
\end{itemize}
In addition to the existing audio files, Switchboard contains highly accurate transcripts and numerous metadata files corresponding to each call and speaker, including speaker sex, age, and dialect. We hope the release of the these labels will encourage others to explore the relationship between speech and emotion in spontaneous speech.

\begin{table*}[t]
    
    \caption{Summary of guidelines provided to graders. These guidelines helped to anchor how graders perceived emotional speech relative to neutral speech, but we emphasized that the graders should rate the speech based on their overall perception.}
    
    \centering
    \begin{subtable}{\textwidth}
    \caption{Guidelines for categorical grading}
    \begin{tabular}{p{10cm}|p{3.2cm}|p{3.2cm}}
        \toprule
         \textbf{Emotion Description} & \textbf{Associated Feelings} & \textbf{Voice Descriptors}\\
         \toprule
         \textbf{Anger} is the feeling of being blocked in our progress or when a personal value or boundary is violated. It also arises in response to perceived provocations or threats. & Rage, Bitterness, Frustration, Annoyance & Harsh, Sharp, Terse \\
         \midrule
         \textbf{Contempt} is the feeling of deep dislike or disrespect toward someone or something you consider unworthy or inferior. & Scorn, Disdain & Smug, Disapproving\\
         \midrule
         \textbf{Disgust} is the feeling of distaste or revulsion toward something that is considered offensive, unpleasant, or gross. & Revulsion, Aversion, Dislike & Negative, Nasal \\
         \midrule
         \textbf{Fear} is a response to a perceived threat or danger, causing feelings of worry or the instinct to protect oneself. & Panic, Anxiety, Nervousness, Overwhelm & Strained, Erratic \\
         \midrule
         \textbf{Sadness} is the feeling of sorrow or unhappiness, often in response to loss, disappointment, or a difficult situation. & Grief, Disappointment & Dull, Flat, Somber \\
         \midrule
         \textbf{Surprise} is triggered by sudden and unexpected information or stimuli from the internal or external environment - like hearing surprising news or having a sudden stomach pain. & Amazement, Distraction & Sudden increase in pitch \\ 
         \midrule
         \textbf{Happiness} is the feeling of pleasure and is often experienced when things are going well or when a person feels fulfilled. & Excitement, Passion, Joy & Positive, Bright \\
         \midrule
         \textbf{Tenderness} is triggered by the feeling of affection, care, and gentle kindness towards oneself, someone or something. This often involves a sense of closeness and compassion. & Compassion, Love, Empathy, Gratitude & Gentle, Friendly \\
         \midrule
         \textbf{Calmness} is the feeling of satisfaction, where one is happy with what they have. & Peaceful, Pleased & Soft \\
         \bottomrule
    \end{tabular} \\
    \end{subtable}\\
    \vspace{8pt}
    \begin{subtable}{\textwidth}
    \caption{Guidelines for dimensional grading}
    \begin{tabular}{p{5cm}|p{5.7cm}|p{5.7cm}}
        \toprule
         \textbf{Dimension Description} & \multicolumn{2}{c}{\textbf{Voice Descriptors}}\\
         \toprule
         \textbf{Dominance} relates to how strong or confident versus weak or uncertain a speaker sounds. & High dominance (confident, strong) may be expressed with direct or concise speech, and perhaps a steady speaking rate or a higher volume. & Low dominance (uncertain, weak) may be expressed with wavering or hesitant speech, and perhaps a choppy speaking rate or lower volume. \\
         \midrule
         \textbf{Valence} relates to how positive versus negative a speaker sounds. & High valence (positive) may be expressed with a pleasant or upbeat tone, and perhaps a higher pitch or speaking rate. & Low valence (negative) may by expressed with a gloomy or harsh tone, and perhaps a lower speaking rate or pitch. \\
         \midrule
         \textbf{Activation} relates to how energized versus drained a speaker sounds. & High activation (energy) may be expressed with a fast (or variable) speaking rate, higher (or variable) volume, or intentional emphasis in speech. & Low activation (drained) may be expressed with a monotone voice, a slow speech rate, or low volume. \\
         \bottomrule
    \end{tabular}
    \end{subtable}\\    
    \label{tab:cues}
    \vspace{-4pt}
\end{table*}

\section{SWB-Affect Annotation}

We selected 10,000 segments from the Switchboard corpus, amounting to roughly 25 hours of speech, and we trained a crowd to annotate the data for categorical and dimensional emotion traits. This section provides background on the Switchboard corpus as well as details of our segment selection, grader training and certification process, and quality analysis.  




\textbf{Switchboard corpus.} Switchboard is a widely analyzed LDC dataset of telephone conversations. We worked with Switchboard-1 Release 2 which contains 260 hours of speech from 543 speakers \cite{godfrey1992switchboard}. Switchboard was collected using a computer-driven robot operator system which paired participants on the phone and prompted them to discuss a specific topic (e.g., care of the elderly, public education, taxes). The data were segmented \cite{deshmukh1998resegmentation}, transcribed, and labeled for echo, static, and background noise. The data were also collected with speaker information, including speaker sex, age, and dialect. The naturalistic conversations in Switchboard have allowed for a range of analyses, including topic classification \cite{mcdonough1994approaches}, dialogue act prediction \cite{colombo2020guiding}, and disfluency detection \cite{romana2023toward, romana2024automatic}. The prompt topics and flow of conversation also have the potential to evoke emotions, which we aimed to explore with the introduction of annotations in Switchboard-Affect.  

\textbf{Segment selection.} We selected segments for annotation (10,000 segments) and a pilot study (100 segments) by first filtering out segments that were more likely to have low quality audio (echo, static, or background noise ratings of 4), not enough content for emotion perception (less than 5 seconds or 5 words), or shifting emotions (greater than 15 seconds). Then, following guidance from the labeling of MSP-Podcast \cite{lotfian2017building}, we reduced the proportion of neutral samples by mining the candidate samples for emotional content. Specifically, we ran an in-house SER model trained for valence, activation, and dominance prediction. We binned the predictions for each dimension into three levels (e.g., low valence, medium valence, high valence) and evenly sampled segments with each of the 27 valence-activation-dominance level combinations.

\textbf{Annotation questions and options.} The annotation tool presented graders with one random audio segment at a time and a series of five questions, similar to the tool used in \cite{lotfian2017building}.

The first two questions pertained to labeling categorical emotions, and asked the grader to select all emotions they perceived followed by the primary emotion they perceived. The options were presented in a list: anger, contempt, disgust, sadness, fear, surprise, happiness, tenderness, calmness, neutral, or other. Note that the first seven emotions in this list come from Ekman's universal set \cite{ekman1992argument} and are commonly used in speech affect labeling \cite{lotfian2017building}. We added tenderness and calmness as options to increase granularity in the positive-valence space. Tenderness promotes prosocial behavior \cite{kalawski2010tenderness}, calmness is linked to self-regulation \cite{hamm2021tale, cordaro2024contentment}, and both have significance in well-being research.  

The next three questions asked the graders to rate the speech for valence, activation, and dominance. For these questions we presented the graders with a 5-point Likert scale. In the valence scale, we indicated that 1=negative, 3=neutral, and 5=positive. In the activation scale, we indicated that 1=drained, 3=neutral, and 5=energized. In the dominance scale, we indicated that 1=weak, 3=neutral, and 5=strong. 

\textbf{Annotation guidelines.} We leveraged a wide body of literature on emotion and its expression in speech to develop a set of training material for this task \cite{ekman1992argument, atlas_of_emotions, grimm2006segregated, andersen1996bright, sobin1999emotion, sauter2009universal}. We presented the material in a self-paced slide deck that included an overview of the task, guidelines for annotating categorical emotions, guidelines for annotating dimensional emotions, and several graded examples with reasoning. 

The overview introduced the task and tool, and it explained a key concept: emotion can be conveyed in both \textit{what} a speaker says (lexical content) and in \textit{how} they say it (paralinguistics). These cues may align and suggest the same emotion, or they may conflict, for example if the speaker is being sarcastic or recalling a story. We emphasized that when they conflict, the graders should make their selections based on \textit{how} the speaker currently sounds in the segment.

The guidelines for annotating categorical emotion included emotion definitions, associated feelings, and voice descriptors. In addition, we manually selected audio examples to serve as examples, and in doing so we ensured examples for each emotion included male and female speakers, as well as low and high intensities of expression. The guidelines for annotating dimensional emotion included definitions with vocal cues and audio examples for the low-, mid-, and high-points of the scale. Table \ref{tab:cues} summarizes the guidelines.

These guidelines helped promote uniform understanding of emotion expression among graders. Specifically, it helped to define and differentiate between related emotions, such as anger, contempt, and disgust. It also clarified that they were expected to label a wide range of intensities, e.g., annoyance and rage are both types of anger, and disappointment and grief are both types of sadness. Ultimately, we found that providing these guidelines helped the graders align their perception of emotion to accepted definitions. At the same time, we emphasized that the guidelines were not all-encompassing and in the end the annotators should make their selections based on their overall perception of the speech. 


\textbf{Gold set creation.} We used a small set of 100 segments to run pilot and internal analyses. The pilot analysis included approximately 60 crowd-sourced graders and the internal analysis included 3 graders from the author list, all of whom independently graded the 100 segments. These analyses allowed our team to iterate on the guidelines and create a gold set. The gold set consisted of 20 segments representing all emotion categories with both male and female speakers. These segments were specifically selected because all internal graders agreed on the primary emotion, suggesting that the speaker was clearly expressing one, \textit{valid} emotion. We assigned \textit{valid} secondary emotions to each of these segments if any of the graders in the pilot analysis selected that emotion. We assigned \textit{valid} dimensional emotion attributes based on the majority response during the internal analysis.   

\textbf{Grader certification.} After reviewing the training material, graders from a selected crowd attempted the gold set. In order to pass, graders had to 
1.) select valid secondary emotions for $>$ 70\% of segments, 2.) select valid primary emotions for $>$ 70\% of segments, and 3.) select valid dimensional attributes for $>$ 70\% segments. The loose thresholds allowed for variation due to subjective, individual perception, while still maintaining some alignment with the definitions and examples in the guidelines. Ultimately, 29 out of 35 graders passed certification, and these graders worked through the annotation set over the course of four weeks. Graders took on average 30 seconds to label one audio segment, and six graders independently labeled each audio segment.


\begin{figure*}[htbp]
\centering
\begin{minipage}[t]{0.54\textwidth}
    \setlength{\abovecaptionskip}{-2pt} 
    \includegraphics[width=\textwidth]{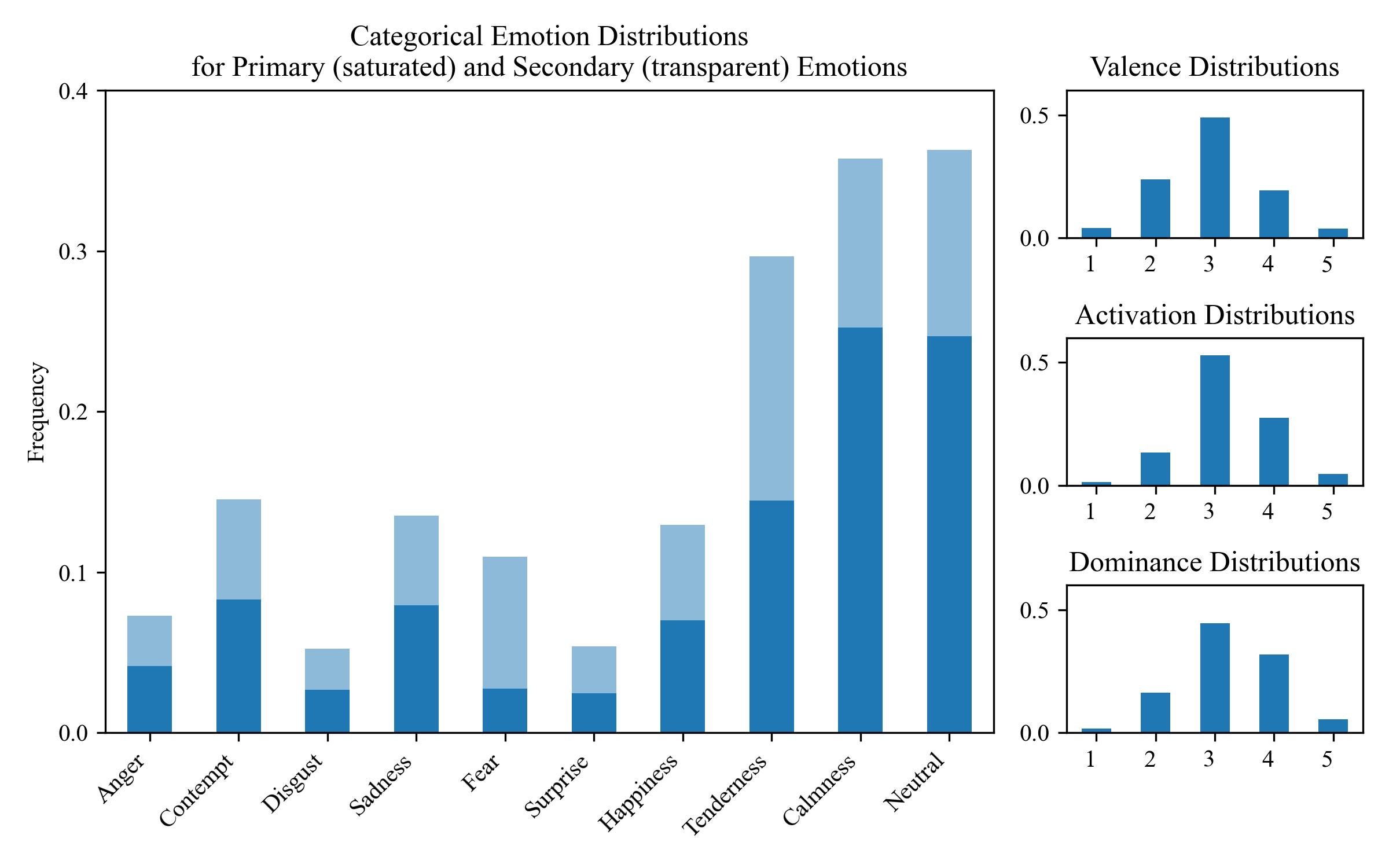}
    \caption{Distribution of affect labels}
    \label{fig:dist}
\end{minipage}
\begin{minipage}[t]{0.44\textwidth}
    \setlength{\abovecaptionskip}{-2pt} 
    \includegraphics[width=\textwidth]{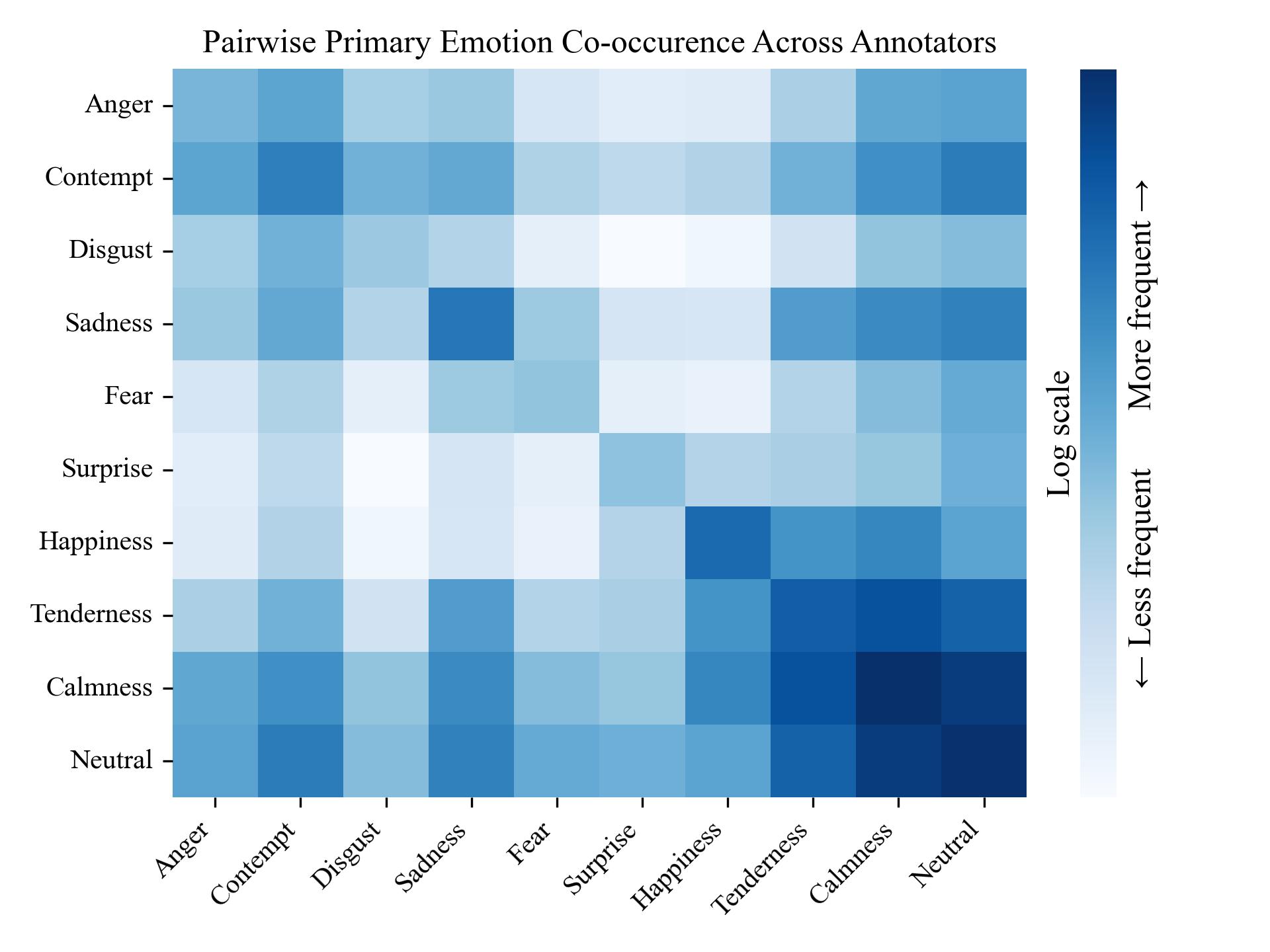}
    \caption{Co-occurence of affect labels}
    \label{fig:cooccurence}
\end{minipage}
\vspace{-4pt}
\end{figure*}

\begin{figure*}
    \centering
    \setlength{\abovecaptionskip}{-0.3pt} 
    \includegraphics[width=\textwidth]{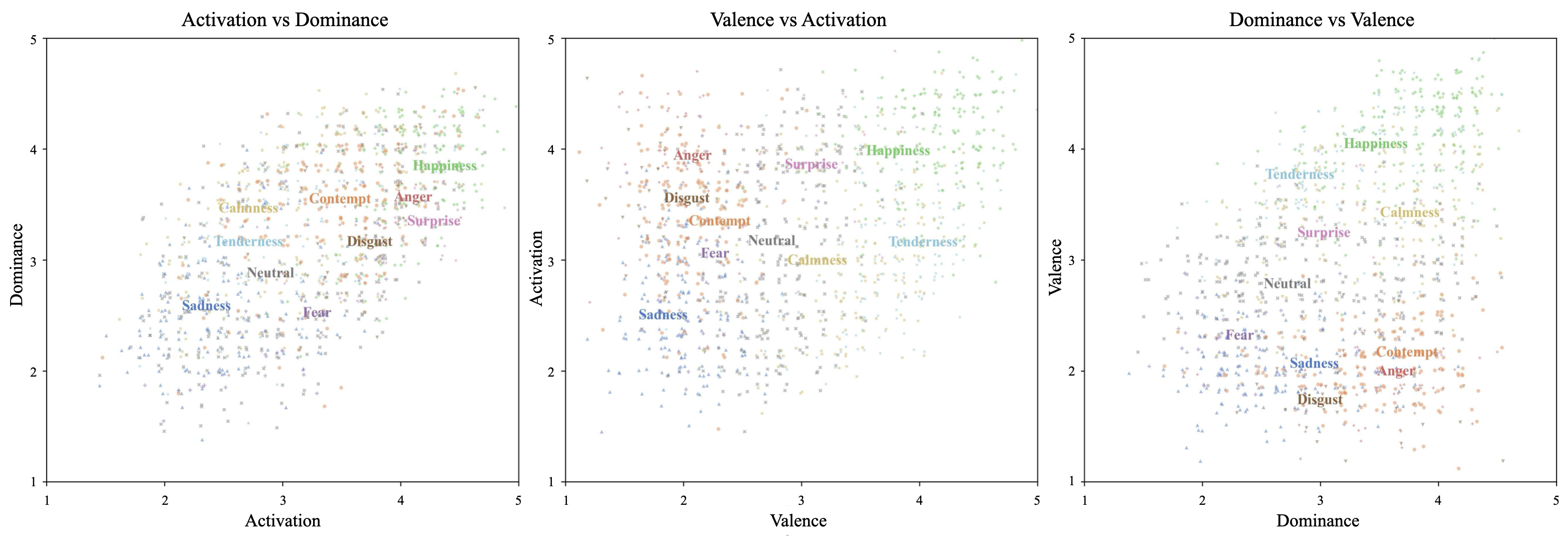}
    \caption{Relationship between dimensional and categorical labels}
    \label{fig:dim_cat}
\vspace{-8pt}
\end{figure*}


\textbf{Affect label Quality Analysis (QA)} Our QA process first focused on grader agreement. For primary emotion and dimensional selections, we calculated Krippendorff's alpha coefficient. For secondary emotion selections, we calculated agreement as the average jaccard similarity of selections between pairs of annotators. We found agreement scores of 0.25 and 0.62 for primary and secondary emotion selection, respectively. We found agreement scores of 0.40, 0.53, and 0.48 for valence, activation, and dominance ratings, respectively. We note emotion grading is notoriously challenging given the complexity of emotional expression in spontaneous speech, and the selection of a primary emotion is especially subjective. Figure \ref{fig:cooccurence} shows the pairwise primary emotion co-occurrence across annotators, illustrating that while agreement for this question was low, annotators were most likely to conflate emotional speech with neutral speech or with closely related emotions (for example contempt and anger). 

In addition, our QA team regularly spot checked annotations and identified several consistent patterns in the labels. First, the emotions skewed toward lower intensity, subtly expressed emotions. This is an expected artifact of the one-on-one conversational data. Secondly, high inter-annotator agreement on a segment tended to indicate more clearly expressed emotions, whereas low agreement corresponded with more ambiguous expressions. Thirdly, for the most part graders appropriately weighted paralinguistic cues, but in some cases graders tagged a segment as emotional when a speaker was recalling an emotional incident instead of actively expressing the emotion. These are ongoing challenges in perceptual grading.  

We release the data in multiple formats, consistent with \cite{lotfian2017building}. We provide consensus labels, where we find consensus primary emotions for 67\% of segments. In addition, we provide detailed annotator-level labels. While the bulk of analysis in this paper focuses on the consensus labels, the detailed labels provide a promising avenue for future work as they may more accurately reflect the nuances in emotion expression and perception, leading to a more precise assessment of SER model performance \cite{tavernor2024whole}. 

\newcolumntype{R}{>{\raggedleft\arraybackslash}Xp{0.01in}}
\newcolumntype{L}{>{\raggedright\arraybackslash}Xp{0.01in}}
\newcolumntype{Y}{>{\centering\arraybackslash}Xp{0.01in}}        

\begin{table*}[h]
\caption{Affect labels in relation to acoustic features. Note: ↑ and ↓ indicate the feature from emotionally-tagged samples is significantly higher or lower, respectively, than the feature from neutral speaker-matched samples in \textcolor[HTML]{2ca02c}{MSP-Podcast Test1}, \textcolor[HTML]{9467bd}{MSP-Podcast Test2}, and \textcolor[HTML]{ff7f0e}{SWB-Affect}. The Wilcoxon test was used to compare the paired data, and significance was assessed with a \textit{p} value $<$ 0.1 after Benjamini-Hochberg correction. MSP-Podcast does not include pause annotations, tenderness labels, or calmness labels. Otherwise, blank cells indicate no significant differences between the emotional and neutral samples.}
\centering
\begin{tabularx}
{\textwidth}
{l X@{\hspace{0.5\tabcolsep}}X@{\hspace{0.5\tabcolsep}}X@{\hspace{5.0\tabcolsep}} 
X@{\hspace{0.5\tabcolsep}}X@{\hspace{0.5\tabcolsep}}X@{\hspace{5.0\tabcolsep}}
X@{\hspace{0.5\tabcolsep}}X@{\hspace{0.5\tabcolsep}}X@{\hspace{5.0\tabcolsep}}
X@{\hspace{0.5\tabcolsep}}X@{\hspace{0.5\tabcolsep}}X@{\hspace{5.0\tabcolsep}}
X@{\hspace{0.5\tabcolsep}}X@{\hspace{0.5\tabcolsep}}X@{\hspace{5.0\tabcolsep}}
X@{\hspace{0.5\tabcolsep}}X@{\hspace{0.5\tabcolsep}}X@{\hspace{5.0\tabcolsep}}
X@{\hspace{0.5\tabcolsep}}X@{\hspace{0.5\tabcolsep}}X@{\hspace{5.0\tabcolsep}}
X@{\hspace{0.5\tabcolsep}}X@{\hspace{0.5\tabcolsep}}X@{\hspace{5.0\tabcolsep}}
X@{\hspace{0.5\tabcolsep}}X@{\hspace{0.5\tabcolsep}}X@{\hspace{5.0\tabcolsep}}}
\toprule
& \multicolumn{3}{l}{Anger} & \multicolumn{3}{l}{\hspace{-0.16in}Contempt} & \multicolumn{3}{l}{\hspace{-0.12in}Disgust} & \multicolumn{3}{l}{\hspace{-0.12in}Sadness} & \multicolumn{3}{l}{\hspace{-0.06in}Fear} & \multicolumn{3}{l}{\hspace{-0.12in}Surprise} & \multicolumn{3}{l}{\hspace{-0.16in}Happiness} & \multicolumn{3}{l}{\hspace{-0.16in}Tenderness} & \multicolumn{3}{l}{\hspace{-0.16in}Calmness} \\

\midrule

Pitch mean & \textcolor[HTML]{2ca02c}{↑} & \textcolor[HTML]{9467bd}{↑} & \textcolor[HTML]{ff7f0e}{↑} & \textcolor[HTML]{2ca02c}{↑} & \textcolor[HTML]{9467bd}{↑} & \textcolor[HTML]{ff7f0e}{↑} & \textcolor[HTML]{2ca02c}{↑} & \textcolor[HTML]{9467bd}{↑} & \textcolor[HTML]{ff7f0e}{} & \textcolor[HTML]{2ca02c}{↓} & \textcolor[HTML]{9467bd}{↓} & \textcolor[HTML]{ff7f0e}{↓} & \textcolor[HTML]{2ca02c}{↑} & \textcolor[HTML]{9467bd}{↑} & \textcolor[HTML]{ff7f0e}{↑} & \textcolor[HTML]{2ca02c}{↑} & \textcolor[HTML]{9467bd}{↑} & \textcolor[HTML]{ff7f0e}{↑} & \textcolor[HTML]{2ca02c}{↑} & \textcolor[HTML]{9467bd}{↑} & \textcolor[HTML]{ff7f0e}{↑} & \textcolor[HTML]{2ca02c}{} & \textcolor[HTML]{9467bd}{} & \textcolor[HTML]{ff7f0e}{} & \textcolor[HTML]{2ca02c}{} & \textcolor[HTML]{9467bd}{} & \textcolor[HTML]{ff7f0e}{↑} \\
Pitch std & \textcolor[HTML]{2ca02c}{↑} & \textcolor[HTML]{9467bd}{↑} & \textcolor[HTML]{ff7f0e}{↓} & \textcolor[HTML]{2ca02c}{↑} & \textcolor[HTML]{9467bd}{↑} & \textcolor[HTML]{ff7f0e}{↓} & \textcolor[HTML]{2ca02c}{↑} & \textcolor[HTML]{9467bd}{↑} & \textcolor[HTML]{ff7f0e}{} & \textcolor[HTML]{2ca02c}{} & \textcolor[HTML]{9467bd}{↓} & \textcolor[HTML]{ff7f0e}{} & \textcolor[HTML]{2ca02c}{↑} & \textcolor[HTML]{9467bd}{↑} & \textcolor[HTML]{ff7f0e}{} & \textcolor[HTML]{2ca02c}{↑} & \textcolor[HTML]{9467bd}{↑} & \textcolor[HTML]{ff7f0e}{↑} & \textcolor[HTML]{2ca02c}{↑} & \textcolor[HTML]{9467bd}{↑} & \textcolor[HTML]{ff7f0e}{↑} & \textcolor[HTML]{2ca02c}{} & \textcolor[HTML]{9467bd}{} & \textcolor[HTML]{ff7f0e}{↑} & \textcolor[HTML]{2ca02c}{} & \textcolor[HTML]{9467bd}{} & \textcolor[HTML]{ff7f0e}{} \\
\midrule 
Loudness mean & \textcolor[HTML]{2ca02c}{↑} & \textcolor[HTML]{9467bd}{↑} & \textcolor[HTML]{ff7f0e}{↑} & \textcolor[HTML]{2ca02c}{↑} & \textcolor[HTML]{9467bd}{↑} & \textcolor[HTML]{ff7f0e}{↑} & \textcolor[HTML]{2ca02c}{↑} & \textcolor[HTML]{9467bd}{} & \textcolor[HTML]{ff7f0e}{↑} & \textcolor[HTML]{2ca02c}{↓} & \textcolor[HTML]{9467bd}{↓} & \textcolor[HTML]{ff7f0e}{↓} & \textcolor[HTML]{2ca02c}{} & \textcolor[HTML]{9467bd}{} & \textcolor[HTML]{ff7f0e}{} & \textcolor[HTML]{2ca02c}{↑} & \textcolor[HTML]{9467bd}{↑} & \textcolor[HTML]{ff7f0e}{↑} & \textcolor[HTML]{2ca02c}{↑} & \textcolor[HTML]{9467bd}{↑} & \textcolor[HTML]{ff7f0e}{↑} & \textcolor[HTML]{2ca02c}{} & \textcolor[HTML]{9467bd}{} & \textcolor[HTML]{ff7f0e}{} & \textcolor[HTML]{2ca02c}{} & \textcolor[HTML]{9467bd}{} & \textcolor[HTML]{ff7f0e}{} \\
Loudness std & \textcolor[HTML]{2ca02c}{} & \textcolor[HTML]{9467bd}{} & \textcolor[HTML]{ff7f0e}{↓} & \textcolor[HTML]{2ca02c}{↓} & \textcolor[HTML]{9467bd}{} & \textcolor[HTML]{ff7f0e}{↓} & \textcolor[HTML]{2ca02c}{} & \textcolor[HTML]{9467bd}{↑} & \textcolor[HTML]{ff7f0e}{} & \textcolor[HTML]{2ca02c}{↑} & \textcolor[HTML]{9467bd}{↑} & \textcolor[HTML]{ff7f0e}{↓} & \textcolor[HTML]{2ca02c}{↓} & \textcolor[HTML]{9467bd}{} & \textcolor[HTML]{ff7f0e}{} & \textcolor[HTML]{2ca02c}{↑} & \textcolor[HTML]{9467bd}{↑} & \textcolor[HTML]{ff7f0e}{↑} & \textcolor[HTML]{2ca02c}{↓} & \textcolor[HTML]{9467bd}{↓} & \textcolor[HTML]{ff7f0e}{} & \textcolor[HTML]{2ca02c}{} & \textcolor[HTML]{9467bd}{} & \textcolor[HTML]{ff7f0e}{↓} & \textcolor[HTML]{2ca02c}{} & \textcolor[HTML]{9467bd}{} & \textcolor[HTML]{ff7f0e}{↓} \\
\midrule
Pause count & \textcolor[HTML]{2ca02c}{} & \textcolor[HTML]{9467bd}{} & \textcolor[HTML]{ff7f0e}{} & \textcolor[HTML]{2ca02c}{} & \textcolor[HTML]{9467bd}{} & \textcolor[HTML]{ff7f0e}{↑} & \textcolor[HTML]{2ca02c}{} & \textcolor[HTML]{9467bd}{} & \textcolor[HTML]{ff7f0e}{↑} & \textcolor[HTML]{2ca02c}{} & \textcolor[HTML]{9467bd}{} & \textcolor[HTML]{ff7f0e}{↑} & \textcolor[HTML]{2ca02c}{} & \textcolor[HTML]{9467bd}{} & \textcolor[HTML]{ff7f0e}{} & \textcolor[HTML]{2ca02c}{} & \textcolor[HTML]{9467bd}{} & \textcolor[HTML]{ff7f0e}{↓} & \textcolor[HTML]{2ca02c}{} & \textcolor[HTML]{9467bd}{} & \textcolor[HTML]{ff7f0e}{↓} & \textcolor[HTML]{2ca02c}{} & \textcolor[HTML]{9467bd}{} & \textcolor[HTML]{ff7f0e}{↑} & \textcolor[HTML]{2ca02c}{} & \textcolor[HTML]{9467bd}{} & \textcolor[HTML]{ff7f0e}{↑} \\
Words per minute & \textcolor[HTML]{2ca02c}{↑} & \textcolor[HTML]{9467bd}{↑} & \textcolor[HTML]{ff7f0e}{↑} & \textcolor[HTML]{2ca02c}{↑} & \textcolor[HTML]{9467bd}{↑} & \textcolor[HTML]{ff7f0e}{↑} & \textcolor[HTML]{2ca02c}{↑} & \textcolor[HTML]{9467bd}{} & \textcolor[HTML]{ff7f0e}{} & \textcolor[HTML]{2ca02c}{↓} & \textcolor[HTML]{9467bd}{↓} & \textcolor[HTML]{ff7f0e}{} & \textcolor[HTML]{2ca02c}{↑} & \textcolor[HTML]{9467bd}{↑} & \textcolor[HTML]{ff7f0e}{} & \textcolor[HTML]{2ca02c}{} & \textcolor[HTML]{9467bd}{↓} & \textcolor[HTML]{ff7f0e}{} & \textcolor[HTML]{2ca02c}{↑} & \textcolor[HTML]{9467bd}{} & \textcolor[HTML]{ff7f0e}{↓} & \textcolor[HTML]{2ca02c}{} & \textcolor[HTML]{9467bd}{} & \textcolor[HTML]{ff7f0e}{↑} & \textcolor[HTML]{2ca02c}{} & \textcolor[HTML]{9467bd}{} & \textcolor[HTML]{ff7f0e}{↑} \\
\midrule
Spectral centroid mean & \textcolor[HTML]{2ca02c}{↑} & \textcolor[HTML]{9467bd}{↑} & \textcolor[HTML]{ff7f0e}{↑} & \textcolor[HTML]{2ca02c}{↑} & \textcolor[HTML]{9467bd}{↓} & \textcolor[HTML]{ff7f0e}{↓} & \textcolor[HTML]{2ca02c}{↑} & \textcolor[HTML]{9467bd}{↑} & \textcolor[HTML]{ff7f0e}{} & \textcolor[HTML]{2ca02c}{↓} & \textcolor[HTML]{9467bd}{↑} & \textcolor[HTML]{ff7f0e}{↓} & \textcolor[HTML]{2ca02c}{↑} & \textcolor[HTML]{9467bd}{↓} & \textcolor[HTML]{ff7f0e}{} & \textcolor[HTML]{2ca02c}{↑} & \textcolor[HTML]{9467bd}{} & \textcolor[HTML]{ff7f0e}{} & \textcolor[HTML]{2ca02c}{↑} & \textcolor[HTML]{9467bd}{↑} & \textcolor[HTML]{ff7f0e}{↑} & \textcolor[HTML]{2ca02c}{} & \textcolor[HTML]{9467bd}{} & \textcolor[HTML]{ff7f0e}{↑} & \textcolor[HTML]{2ca02c}{} & \textcolor[HTML]{9467bd}{} & \textcolor[HTML]{ff7f0e}{} \\
Spectral centroid std & \textcolor[HTML]{2ca02c}{↓} & \textcolor[HTML]{9467bd}{↓} & \textcolor[HTML]{ff7f0e}{} & \textcolor[HTML]{2ca02c}{↑} & \textcolor[HTML]{9467bd}{↓} & \textcolor[HTML]{ff7f0e}{↓} & \textcolor[HTML]{2ca02c}{↑} & \textcolor[HTML]{9467bd}{} & \textcolor[HTML]{ff7f0e}{} & \textcolor[HTML]{2ca02c}{} & \textcolor[HTML]{9467bd}{↑} & \textcolor[HTML]{ff7f0e}{↑} & \textcolor[HTML]{2ca02c}{↓} & \textcolor[HTML]{9467bd}{↓} & \textcolor[HTML]{ff7f0e}{↓} & \textcolor[HTML]{2ca02c}{↓} & \textcolor[HTML]{9467bd}{↓} & \textcolor[HTML]{ff7f0e}{} & \textcolor[HTML]{2ca02c}{↓} & \textcolor[HTML]{9467bd}{↓} & \textcolor[HTML]{ff7f0e}{↓} & \textcolor[HTML]{2ca02c}{} & \textcolor[HTML]{9467bd}{} & \textcolor[HTML]{ff7f0e}{} & \textcolor[HTML]{2ca02c}{} & \textcolor[HTML]{9467bd}{} & \textcolor[HTML]{ff7f0e}{↓} \\
\bottomrule
\end{tabularx}
\label{tab:paralinguistics}
\vspace{-4pt}
\end{table*}

\begin{table}[]
    \caption{GPT-4o prompt, adapted from \cite{niu2024rethinking}, to estimate the probabilities of detecting each emotion from lexical content.}
    \centering
    \begin{tabular}{|p{7.5cm}|}
    \toprule
         \textbf{System} \\
         You are an emotionally-intelligent and empathetic agent. 
         You will be given a transcript from a speaker, and your task is to identify the primary emotion the speaker is expressing within the text. If there is no emotion, then the primary emotion is neutral. 
         Classify the transcript into one of the following categories: anger, contempt, disgust, sadness, fear, surprise, happiness, neutral.
         Respond with only one category and keep your responses to the category name as written and nothing else.  \\ \\
         \textbf{User}\\
         Transcript: $<$segment transcript$>$ \\ \\
         \textbf{Assistant} \\
         The primary emotion is \\
    \bottomrule
    \end{tabular}
    \label{tab:prompt}
\vspace{4pt}
\end{table}

\begin{table}[]
\caption{Affect labels in relation to lexical content. Reported values are average emotion detection probabilities from GPT-4o after prompting with speech transcripts.}
\centering
\begin{tabular}{lc}
\hline
& Switchboard-Affect\\ 
\hline
Anger &0.301\\
Contempt &0.248\\ 
Disgust &0.324\\ 
Sadness &0.419\\ 
Fear &0.586\\ 
Surprise &0.299\\
Happiness &0.174\\ 
Tenderness &0.082\\ 
Calmness &0.048\\ 
\hline
Unweighted Average &0.276\\
Weighted Average &0.193\\ 
\hline
\end{tabular}

\label{tab:lexical}
\vspace{-8pt}

\end{table}

\section{SWB-Affect Analysis}
 
\subsection{Data distributions.} 

Figure \ref{fig:dist} illustrates the distribution of labels in SWB-Affect. We conduct chi-square tests of independence to understand the impact of speaker demographics, namely sex and age, in the emotion labels. We find statistically significant (\textit{p} value $<$ 0.05) associations between sex and emotion labels, with males having more samples labeled as calmness or neutral and females having more samples labeled as happiness or tenderness. In addition, we find statistically significant (\textit{p} value $<$ 0.05) associations between age and emotion labels, with younger individuals (20-30 year olds) having more samples labeled as calmness and older individuals (40+ year olds) having more samples labeled as happiness. These relationships may result from the underlying data collection protocol, or they may reflect biases in the data mining procedure, training material, or graders themselves. Ultimately, these findings underscore the importance of analyzing SER model performance by emotion type and demographic group. 


We also conduct an analysis of the relationship between dimensional and categorical annotations, illustrated in Figure \ref{fig:dim_cat}. We find many expected relationships, for example, happiness is associated with high valence, activation, and dominance, whereas calmness is linked to slightly lower valence, noticeably lower activation, but similar dominance. Sadness and fear are both associated with low valence and low dominance, and fear is related to higher activation. Anger, contempt, and disgust cluster nearby but with key differences. Anger and contempt are associated with higher dominance compared to disgust, and anger is associated with higher activation compared to both contempt and disgust. Although we did not explicitly provide graders with information connecting the categorical and dimensional labeling schemes, the findings are closely aligned with emotion definitions in the literature which highlights the graders strong understanding of the task.

\subsection{Investigating Factors in Speech Emotion Perception}

Understanding the cues that graders rely on when labeling is important to ensure that SER models evaluated with the data are being assessed based on their ability to learn a holistic and accurate representation of emotion perception. In this section, we present an approach to quantify the influence of lexical and paralinguistic factors in the labeling of SWB-Affect. 

\textbf{Lexical factors}. We use the publicly available OpenAI GPT-4o model to understand lexical importance in emotion perception. Previously, Niu et al. have shown that GPT-4 excels at emotion recognition from text, even outperforming human graders at this task\cite{niu2024text, niu2024rethinking}. In this work, we use text-based GPT-4o emotion recognition probabilities to quantify the extent to which human graders may have selected emotion tags based on the lexical content. We modify the prompt from \cite{niu2024rethinking} to have GPT-4o label emotional transcripts in SWB-Affect; our version of the prompt is in Table \ref{tab:prompt}. Note that the latest API does not support extracting specific probabilities. As a workaround, we prompt the model ten times with temperature=1 to encourage varying responses. We take the mean probability associated with the human-labeled consensus emotion. If, within those multiple requests, GPT does not respond with the human-labeled consensus emotion, we assume the probability is 0.  

Table \ref{tab:lexical} lists the text-based emotion detection probabilities from GPT-4o. We find unweighted and weighted average probabilities of 0.276 and 0.193, respectively. The weighted average suffers due to lower average probability of detecting calmness, tenderness, and happiness from text alone (averages of 0.048, 0.082, and 0.174, respectively). This suggests paralinguistic cues may have played a larger role when annotators were distinguishing these emotions from neutral speech. Most other emotions are detected from text with an average probability range of 0.3-0.4. However, fear samples are detected with a relatively high probability (average=0.586) suggesting that the lexical content of the segments may have been more relevant to annotators' perception of fear. Overall the low probabilities of emotion detection from text alone, even with a state-of-the-art GPT-4o model, underscore the importance of acoustic modeling in SER. 

\textbf{Paralinguistic factors}. Previous work has found numerous prosodic and spectral features relevant for distinguishing between neutral and emotional speech, as well as between emotion types \cite{sauter2009universal, schuller2009prosodic, eyben2015geneva}. In this work, we extract and compare acoustic features from speaker-matched emotional and neutral speech segments. Specifically we select a small number of commonly explored, interpretable features: 
\begin{itemize}
    \item Pitch: log F0, mean and standard deviation 
    \item Loudness: log energy, mean and standard deviation
    \item Rhythm: number of pauses and words per minute 
    \item Spectral centroid: mean and standard deviation
\end{itemize}

\noindent We extract the features using the corpus audio, transcripts, and Librosa \cite{mcfee2015librosa}. To reduce the influence of specific matches, we select up to 10 neutral speaker-matched samples for comparison to each emotional sample. We conduct the Wilcoxon test with two alternatives to evaluate which, if any, of these features can be used to distinguish between emotional and neutral speech in the dataset. We use the Benjamini-Hochberg correction to adjust \textit{p} values. We also repeat the analysis with MSP-Podcast Test1 and Test2 sets to validate our findings, although MSP-Podcast does not include pause annotations, tenderness labels, or calmness labels. 



Table \ref{tab:paralinguistics} illustrates our findings on the relevant paralinguistic cues. We find several consistent trends across all three datasets, particularly with respect to pitch and loudness. While we find some disagreements, the within-corpus differences between MSP-Podcast test sets suggest that some variation is expected due to variability in emotion expression, emotion perception, and the broad nature of these emotion categories. 


Overall, we find acoustic features in emotionally-tagged segments in SWB-Affect align with previous work in this space. For example, samples tagged as happy or surprise have a higher pitch and loudness, whereas samples tagged as sad have a lower pitch and loudness \cite{murray1993toward}. Furthermore, samples tagged as angry have a higher spectral centroid mean compared to samples tagged as sad \cite{wu2011automatic}. We also observe that the relevant cues found for anger, contempt, and disgust reveal some commonalities and differences between the often confused emotions. All three correspond to higher loudness, but only anger and contempt are associated with higher pitch and higher speaking rates compared to neutral speech. On the other hand, contempt and disgust co-occur with more pauses compared to neutral speech. Lastly, anger and contempt are linked to higher and lower spectral centroid means, respectively, compared to neutral speech. This analysis identifies key paralinguistic cues that are relevant to capture in SER modeling, particularly for naturalistic conversational speech.

\begin{table*}[h]
    \caption{SER model F1 and recall scores for primary emotion classification in SWB-Affect. 
    }
\centering
    \setlength{\abovecaptionskip}{-4pt} 

    \centering
    \begin{tabular}{lccccccccc}
    \hline
    F1-Score & Anger & Contempt & Disgust & Sadness & Fear & Surprise & Happiness & Neutral & Unweighted Average \\ \hline 
    Emotion2Vec & 0.000 & -- & 0.083 & 0.420 & 0.240 & 0.021 & 0.710 & 0.708 & -- \\
    Odyssey & 0.163 & \textbf{0.153} & 0.173 & 0.404 & 0.065 & \textbf{0.244} & 0.739 & 0.440 & 0.300 \\ 
    Whisper-GRU & 0.147 & 0.060 & 0.024 & 0.453 & 0.089 & 0.211 & \textbf{0.844} & \textbf{0.778} & 0.326 \\
    GPT-4o & \textbf{0.257} & 0.077 & \textbf{0.222} & \textbf{0.606} & \textbf{0.364} & 0.241 & 0.589 & 0.773 & \textbf{0.391} \\ 
    \hline \\
    \hline 
    Recall Score & Anger & Contempt & Disgust & Sadness & Fear & Surprise & Happiness & Neutral & Unweighted Average \\ \hline 
    Emotion2Vec & 0.000 & -- & 0.051 & 0.395 & 0.141 & 0.011 & 0.679 & 0.776 & -- \\
    Odyssey & 0.096 & \textbf{0.094} & \textbf{0.141} & \textbf{0.937} & 0.129 & \textbf{0.370} & \textbf{0.861} & 0.291 & \textbf{0.365} \\ 
    Whisper-GRU & 0.088 & 0.032 & 0.013 & 0.420 & 0.047 & 0.163 & 0.854 & 0.939 & 0.319 \\
    GPT-4o & \textbf{0.325} & 0.041 & 0.128 & 0.605 & \textbf{0.236} & 0.152 & 0.430 & \textbf{0.951} & 0.358 \\
    \hline \\ 
    \end{tabular}

    \label{tab:catresults}
\vspace{-12pt}

\end{table*}

\begin{table}[h]
    \caption{SER model CCC for dimensional attribute prediction in SWB-Affect.
    }

\centering
    \centering
    \begin{tabular}{lccc}
    \hline
                & Activation & Valence & Dominance \\ \hline
    Audeering W2V2 & \textbf{0.455} & 0.648 & \textbf{0.424}\\ 
    Odyssey & 0.328 & \textbf{0.689} & 0.240 \\ 
    Whisper-GRU & 0.400 & 0.666 & 0.300\\
    GPT-4o & 0.210 & 0.674 & 0.384\\ 
    \hline
    \end{tabular}

    \label{tab:dimresults}
\vspace{-12pt}
\end{table}

\subsection{Speech Emotion Recognition Benchmarking}

We evaluate existing categorical and dimensional SER models on SWB-Affect to understand their generalizability to spontaneous, conversational speech. We evaluate the categorical SER models on all segments that have a consensus label of neutral or one of Ekman's seven universal emotions (anger, contempt, disgust, sadness, fear, surprise, happiness) as these are the most common output categories for existing models. We evaluate the dimensional SER models on all segments. Before running the evaluations, we upsample the audio from 8kHz to 16kHz.

\textbf{SER models.} We compare performance from four previously published SER models, and one zero-shot approach: 
\begin{itemize}
    \item Emotion2Vec \cite{ma2023emotion2vec}. A universal speech representation model pre-trained with unlabeled emotional data and fine-tuned with IEMOCAP categorical labels. 
    \item Audeering W2V2 \cite{wagner2023dawn}. A model based on wav2vec 2.0 (W2V2, \cite{baevski2020wav2vec}), which was pruned to 12 transformer layers and fine-tuned with MSP-Podcast dimensional labels. 
    \item Odyssey \cite{Goncalves_2024}. A model based on WavLM \cite{chen2022wavlm}, which has been fine-tuned with MSP-Podcast consensus categorical and dimensional labels. 
    \item Whisper-GRU \cite{mitramodeling}. A model that uses frozen Whisper embeddings \cite{radford2023robust} as input to a GRU network trained with MSP-Podcast categorical soft labels (distributions of perception, not consensus labels) and dimensional consensus labels.
    \item GPT-4o (audio preview) \cite{openai_docs}. A GPT-based model that accepts audio inputs and prompts but has not been trained with speech emotion targets. When prompting GPT-4o to select an emotion given an audio sample, we use a prompt similar to that in Table \ref{tab:prompt}.
\end{itemize} 

Note that Emotion2Vec only provides categorical predictions (and does not include contempt), Audeering only provides dimensional predictions, and Odyssey, Whisper-GRU, and GPT-4o provide both. 

\textbf{Metrics.} We present results separately for categorical and dimensional SER models. For the former, we report F1-score and recall for primary emotion prediction at the class-level as well as unweighted averages. 
For dimensional SER performance, we report Lin's Concordance correlation coefficient (CCC) between predicted and ground truth labels. 

\textbf{Results.} Table \ref{tab:catresults} lists the categorical emotion prediction results by model. We find GPT-4o performs best in terms of unweighted average F1-score (0.391 vs 0.300-0.326) and significantly outperforms other models for most classes (e.g., anger, sadness, fear) while underperforming for happiness. Odyssey performs best in terms of unweighted average recall (0.365 vs 0.319-0.358), and the recall of most classes. However, its recall of neutral, the majority class, is significantly lower than with other models (0.291 vs 0.776-0.951). 

For most models, fear is one of the hardest emotions to detect while happiness is the easiest, which mirrors the class frequencies, but this performance trend is the opposite of our findings with lexical-based emotion predictions. This suggests the current models would benefit by increasing fear representation during training, or alternatively, by considering lexical content more explicitly when classifying fear. We also find lower probabilities for surprise, which we find is often misclassified as happiness, potentially due to the paralinguistic similarities between the two emotions. Similarly, we find confusion between anger, contempt, and disgust, but future work may benefit from training or prompting models with distinguishing paralinguistic cues in mind. For example, Perez et al. have previously demonstrated that pause information can improve lexical-based SER \cite{perez2022mind}, and our analysis suggests the potential of additional prosody or spectral features in this type of multimodal framework.

Table \ref{tab:dimresults} lists the dimensional emotion prediction results by model and attribute. We find Odyssey performs best for valence prediction (CCC=0.689) but other models perform similarly (0.648-0.674). Audeering performs best for activation and dominance prediction (CCC=0.455 and 0.424, respectively) and other models vary considerably in their performance (0.210-0.400). 

When comparing these results to results on MSP-Podcast Test1 and Test2, we find a few consistent trends. First, all models detect sadness and happiness with higher accuracy in SWB-Affect compared to MSP-Podcast, but at the same time, all models detect anger with lower accuracy in SWB-Affect compared to MSP-Podcast. We suspect this is related to how speakers express each emotion in different contexts. For example, speakers may be more overtly expressive of sadness or happiness in one-on-one conversations with strangers, but more overtly expressive of anger in podcasts. Second, we find that all models predict valence comparably across SWB-Affect and MSP-Podcast, but that they predict activation and dominance with lower accuracy in SWB-Affect compared to MSP-Podcast. We attribute this again to the change in domain where podcast speech likely has different activation and dominance qualities given that speakers are presenting to an audience. As a result, MSP-Podcast may have been labeled with different representations of these dimensions in mind. These findings underscore the need to evaluate SER models with speech from a variety of contexts.


\section{Conclusion}

This work introduces a new set of emotion labels for the existing Switchboard corpus. The Switchboard corpus includes speech segments from naturalistic conversations and it has been widely transcribed and annotated. In this work, we describe our method for training a crowd to label emotion perception in Switchboard, and we explore trends in the annotated data. We find low agreement for primary emotion selection, but that the disagreements may meaningfully point to mixed or ambiguous emotions. Future work may benefit from evaluation with individual annotator labels or distributions. In addition, we investigate the lexical and paralinguistic cues that graders may have perceived when labeling the data. We find the labels for fear are linked to lexical content, whereas the labels for happiness correspond more to paralinguistic features. Future work may benefit from considering both text and audio-based inputs to capture the complexity of emotion expression. Lastly, we explore the performance of state-of-the-art SER models on the newly labeled data. While SER performance drops for anger, it increases for happiness and sadness in SWB-Affect. This likely points to the variability of emotion expression in different domains and underscores the need for diverse evaluation datasets. Ultimately, the affective labeling of Switchboard opens many future analysis directions, including improving SER for low-intensity, conversational emotions, as well as analyses on the relationships between emotion perception and turn-taking, dialogue acts, or speech disfluencies. 

\section{Ethical Impact Statement}

SWB-Affect contributes to the evaluation of SER models, which may be used in a range of applications including healthcare, customer satisfaction, entertainment, or education. Any assessments of SER model performance on SWB-Affect may be influenced by underlying biases in the data. Perceptual emotion annotation is a difficult, subjective task. In addition, emotion expression varies from person to person and can be especially subtle in conversational speech. The speakers in this task are US-based English speakers, and the findings may not generalize to other languages or accents. In addition, the graders employed in this task are US-based English speakers, and the labels may not generalize to perceptions of emotion that are common in other languages or cultures. Lastly, the labels may be influenced by insufficient context, individual annotator biases, or biases that shaped the overall annotation process. Specifically, biases may have played a role in segment selection, in the simplified emotion definitions, in the voice descriptors and audio examples, or in the gold set we used to certify annotators. For transparency, we provide detailed summaries of our training guidelines and processes. Future annotation work should include a larger and more diverse group of annotators to minimize risk of individual biases. In addition, future annotation work should explore continuous grading with context as has been done recently in other datasets \cite{kossaifi2019sewa, martinez2020msp, poria2018meld}. 

We also acknowledge the limitations of the analysis section. The investigation on lexical and paralinguistic cues provides a post-hoc analysis of the different factors that may have played a role in annotators' perception of emotion. The list of cues analyzed is not exhaustive, and future work may improve on this process by collecting annotator reasoning during the labeling. In addition, our evaluation of state-of-the-art SER models may by limited by audio quality or label reliability, as mentioned above. However, the overall trends we see across all models highlight the importance of evaluation with datasets that capture natural affective variations in speech. 

\bibliographystyle{IEEEtran}
\bibliography{references}

\vspace{12pt}

\end{document}